\documentclass[prapplied, amsmath,amssymbn,twocolumn]{revtex4-1}

\usepackage{graphicx}
\usepackage{color}

\begin{document}

\title{Alloy, Janus and core-shell nanoparticles: Numerical modeling\\
 of their nucleation and growth in physical synthesis}

\author{Georg Daniel F{\"o}rster}
\affiliation{Laboratoire d'\'Etude des Microstructures, ONERA-CNRS, UMR104, Universit\'e Paris-Saclay, BP 72, 92322 Ch\^atillon Cedex, France}
\author{Magali Benoit}
\affiliation{CEMES, Universit\'e de Toulouse, CNRS, 29 rue Jeanne Marvig,Toulouse 31055, France}
\author{Julien Lam}
\email{julien.lam@ulb.ac.be}
\affiliation{Center for Nonlinear Phenomena and Complex Systems, Code Postal 231, Universit\'e Libre de Bruxelles, Boulevard du Triomphe, 1050 Brussels, Belgium}

\begin{abstract}
While alloy, core-shell and Janus binary nanoclusters are found in more and more technological applications, their formation mechanisms are still poorly understood, especially during synthesis methods involving physical approaches. In this work, we employ a very simple model of such complex systems using Lennard-Jones interactions and inert gas quenching. After demonstrating the ability of the model to well reproduce the formation of alloy, core-shell or Janus nanoparticles, we studied their temporal evolution from the gas via droplets to nanocrystalline particles. In particular, we showed that the growth mechanisms exhibit qualitative differences between these three chemical orderings. Then, we determined how the quenching rate can be used to finely tune structural characteristics of the final nanoparticles, including size, shape and crystallinity.
\end{abstract}
\maketitle

\section{Introduction}

Envisioned applications of nanoparticles (NPs), including catalysis\cite{Walker2016Jan,Ho2014May,Singh2017Aug,Song2018Dec,Weng2018Apr}, optics\cite{Yang2018Dec,LopezOrtega2018Oct,Dixit2018Feb,Makarov2018Jun,Fenwick2016Jun,Liang2019May}, and medicine\cite{Wang2019May,Tan2018Sep,Xu2019May,Liu2019May,Gil2018Nov}, depend on the ability to tailor their physical properties. A precise control of composition, shape, size distribution, and crystal structure should therefore be achieved during the synthesis~\cite{Zhao2013Feb,Dhand2015Dec,Albanese2012Jul}. Most recent research efforts went into raising the complexity of the material and multi-component systems are increasingly researched in order to combine several properties within the same NP\cite{Yan2017Mar,Chaudhuri2011Dec,Zhang2017Jul}. For binary systems, three different types of morphology are observed (1) alloy\cite{Reineck2019Mar,Cui2013Jun,Zhang2018Oct2,Lucci2015Oct}, (2) core-shell\cite{Langlois2015Jul,Ramade2017Sep,Fu2019Apr,Manigandan2019Apr,Li2017Feb,Amendola2019Jul} and (3) Janus\cite{Liang2019May,Zhou2019Mar,Liu2018Dec,Ju2017Sep}. 

Owing to their simplicity and their ability to generate a wealth of structures, physical synthesis methods represent an appealing alternative to more traditional soft chemistry routes and should be crucial to reach an on-demand synthesis of nanomaterials in general\cite{Naatz2017Jan,Wegner2018Sep,Amendola2019Jul,LiYang2006Sep,Langlois2015Jul,Zhang2018Oct,Zhao2016Mar,Srivastava2014Nov,Wu2014Feb,Levi2010Jul,Wu2011Jan}. Physical synthesis methods include laser ablation~\cite{Ayyub2001Jul,Semaltianos2010May,Zhang2017Feb,Kim2017}, flame pyrolysis~\cite{Rosnerc2005Jul,Teoh2010Aug,Li2016Jul}, or magnetron sputtering~\cite{Ayyub2001Jul,Kelly2000Mar,Alexeeva2016Feb} and always start with the production of an initially hot gas. The subsequent cooling renders the system unstable, which leads to the nucleation and growth of the NPs. These methods are highly versatile and easily controllable via experimental parameters. The out-of-equilibrium state during the growth process and the formation kinetics both play a crucial role in determining particle properties\cite{Kathmann2009Jun,Stankic2011Aug} and should therefore help obtaining new exotic structures and morphologies~\cite{Amendola2019Jul,Levi2010Jul,LiYang2006Sep,Langlois2015Jul,Zhang2018Oct}.

However, due to their complexity, details of the NP formation processes are still not very well understood. Experimentally, physical synthesis has been investigated mostly by optical techniques~\cite{Li2016Jul}. In particular, at the early stages, when gas and small clusters make up most of the system, the temperature can be probed by infrared~\cite{Rudin2012Jun,Grohn2012May} and Raman spectroscopy~\cite{Liu2010Sep,Zhang2013Mar}, laser-induced fluorescence~\cite{Colibaba2000Aug}, and laser-induced breakdown spectroscopy~\cite{Lam2013Dec,Zhang2013Mar,Amodeo2008Oct}. For instance, with the latter technique, the chemical composition of diatomic molecules and the electron density were recently measured for laser ablation in air and liquid~\cite{Lam2014Nov,Lam2015Sep,Amans2017Mar,Lam2013Dec}. The synthesized nanomaterials can also be characterized in terms of size and crystallinity using scattering methods based on optical or X-ray sources~\cite{Forster2015Dec,Ibrahimkutty2012Sep,Kousal2018Oct,Reich2019}. Understanding the mechanisms occurring in physical synthesis techniques has also benefited from numerical simulations\cite{Shih2018Apr}. For instance, models for laser-matter interaction and the subsequent plasma formation have been developed for the study of laser ablation~\cite{Forster2018Jun}. 

Nevertheless, the growth of nanostructures in the cooling ablation plume appears to be much harder to study. On the one hand, nucleation and growth in such non-equilibrium processes are very fast and involve few atoms, so that it cannot be easily investigated using conventional experimental techniques. On the other hand, much numerical effort went into the investigation of this problem and gas condensation in general. The temperature quench is usually achieved by successive collisions between the reactive and inert gas atoms kept at room temperature. With such modeling, it was demonstrated that the condensation rate is almost independent of the interatomic potential and that it can be related to the partial pressure of the inert gas using an analytical model~\cite{Kesala2007May}. Various monoatomic materials were studied, including silicon~\cite{Zhao2015Jan}, germanium~\cite{Krasnochtchekov2005Jan,Krasnochtchekov2005Oct}, and metals~\cite{Kesala2007May} using semi-empirical potentials such as EAM, Tersoff, and Stillinger-Webber. Yet, for binary systems like alloy, Janus and core-shell nanomaterials, the atomistic mechanisms occurring during the synthesis are not very clear yet, especially in the context of physical methods~\cite{parsina2010tailoring,singh2014heterogeneous,Nelli2019Jul}. 

In this work, we study three binary Lennard-Jones systems each tailored to favor one of these morphologies. On the one hand, we identify the differences in the synthesis mechanisms for each morphology by studying the temporal evolution. On the other hand, the final distributions of clusters are analyzed in terms of structural properties, including size, asphericity, crystallinity and, chemical ordering. Moreover, we show that these structural properties can be finely adjusted solely by changing the quenching rate, which can be controlled in most experimental set-ups. Using this simple model system, we aim at drawing very general conclusions regarding the underlying mechanisms which involve both thermodynamical and kinetic processes.

\section{Methods}

Three different types of atoms make up the systems studied here. Truncated and shifted Lennard-Jones interactions are used throughout this work. The two first species are reactive, and their interactions are tailored to obtain (1) alloy, (2) Janus and (3) core-shell NPs [See Table\,I]. In the two latter cases, the coefficients are chosen by following the work of Mravlak et al. who computed the most stable structures of binary nanoclusters for a large number of combination of coefficients~\cite{Mravlak2016Jul}. In the alloy case, the $\sigma$ parameters are chosen to obtain the Kobb-Andersen system, which crystallizes in alloy BCC structures~\cite{Kob1994Sep,Kob1995May,Kob1995Oct,Fernandez2003Jan}, while the energy parameters are set in a way to roughly match the melting and boiling temperatures obtained with the core-shell and Janus systems. The third species serves as an inert buffer gas and interacts with the reactive NPs and with one another also using truncated and shifted Lennard-Jones interactions, but with a cutoff equal to $2{^{(1/6)}}\sigma$ which leads to a purely repulsive interaction. 

\begin{table*}[]
\begin{center}
\begin{tabular}{l|c|c|c||c|c|c||c|c|c|}
\cline{2-10}
                          & \multicolumn{3}{c||}{Alloy}         & \multicolumn{3}{c||}{Janus}  & \multicolumn{3}{c|}{Core-shell}        \\ \hline
\multicolumn{1}{|l|}{}    & $\epsilon_{ij}$   & $\sigma_{ij}$ & $r_{cut} (\sigma)$ & $\epsilon_{ij}$ & $\sigma_{ij}$ & $r_{cut} (\sigma)$ & $\epsilon_{ij}$ & $\sigma_{ij}$ & $ r_{cut} (\sigma)$ \\ \hline 
\multicolumn{1}{|l|}{A-A} &         0.80 &  1.00    &    2.5   &  1.00 &  1.00    &    2.5 &            1.00 &  1.00    &    2.5         \\ \hline
\multicolumn{1}{|l|}{B-B} &         0.40 &  0.88    &    2.2   &  0.62 &  1.00    &    2.5 &            0.50 &  1.00    &    2.5         \\ \hline
\multicolumn{1}{|l|}{A-B} &         1.20 &  0.80    &    2.0   &  0.30 &  1.00    &    2.5 &            0.70 &  1.00    &    2.5         \\ \hline
\end{tabular}
\caption{Lennard-Jones parameters for the interaction between reactive NPs in the three different cases.}

\end{center}\end{table*}

\begin{figure}[h!]
\includegraphics[width=\columnwidth]{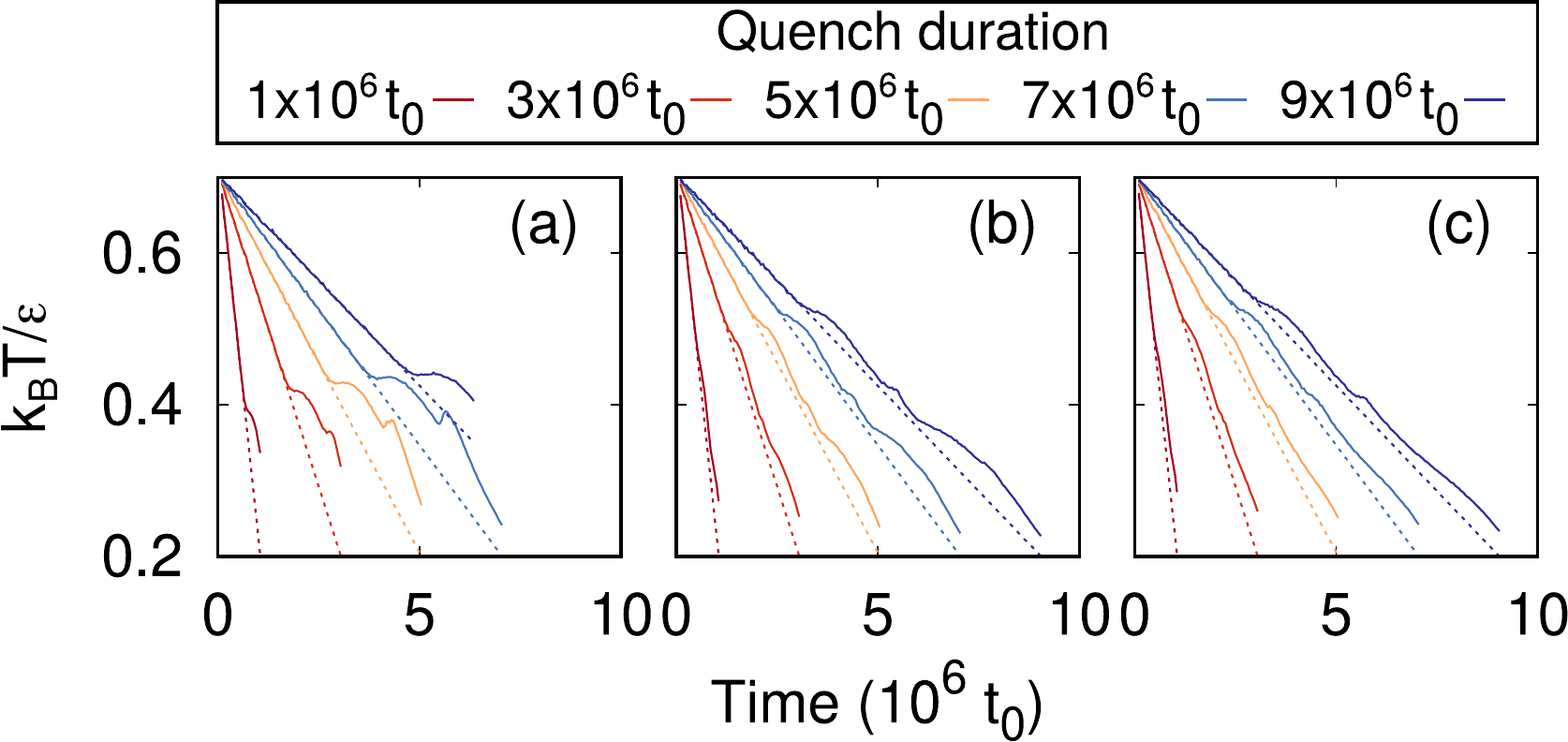}
\caption{Temperature evolution during the quench in the (a) alloy, (b) Janus and (c) core-shell cases. Continuous and dotted lines represent the temperature of the reactive and inert gas, respectively.}
\label{Temperature}
\end{figure}

Initially, the reactive system consists of $10^5$ atoms at a density of 0.02$\sigma^{-3}$. 
The same number of inert atoms is added in the simulation box. In our molecular dynamics simulations, the reactive atoms move freely (non-thermostated), a Berendsen thermostat controls the temperature of the inert gas. The quenching is obtained by linearly decreasing the inert gas temperature and thus inducing a subsequent cooling of the reactive gas. The temperature is decreased from $k_BT = 0.7\epsilon$ to $k_BT = 0.2\epsilon$ during quench durations $\Delta t$, ranging from $10^6t_0$ to $10^7t_0$. The simulation time step $t_0$ correpsonds to $0.005 \tau$ in Lennard-Jones units. To convert Lennard-Jones reduced units into physical units, one can use Lennard-Jones parameters obtained for aluminium where $\tau=0.19$\,ps and $\epsilon/k_B=5886$\,K~\cite{Filippova2015Jan}. In this case, the corresponding quenching rates range from $3080$~K/ns to $308$~K/ns which is faster than experimental rates measured in the case of plasma quenching with laser ablation~\cite{Amans2017Mar}. This simulation protocol allows for a good control of the quenching rate without having to finely tune neither the inert gas pressure nor its interactions with the reactive atoms. Fig.\,\ref{Temperature}, shows that initially, the reactive gas temperature follows closely the imposed  constant quenching rate. Then, because of condensation and crystallization, potential energy is released that adds to the kinetic energy. This trend was also observed when instead of a linear decrease, the inert gas temperature was fixed at low temperatures throughout the entire simulations~\cite{Kesala2007May}. Final results are averaged over ten independent runs. All the simulations were performed using the LAMMPS software package~\cite{Plimpton1995Mar}.

For the analysis, atoms less than 1.5$\sigma$ apart from each other are regrouped into clusters. In what follows, any clusters made of more than one atom is referred to as ``NP". For each of the obtained NPs, several characterizations are carried out. The size is measured simply by counting the number of atoms and by means of the gyration radius of the NPs, i.e., the root mean square distances of the atoms from the center of mass of the NPs. Furthermore, we quantify the shape of the NPs with the help of an asphericity index, $\chi=3\left(2I_1-I_2-I_3\right)/2\left( I_1+I_2+I_3\right)$, with $I_1\geq I_2\geq I_3$ the principal moments of inertia of the NP. The asphericity index varies by construction between 0 and 1, with values close to zero corresponding to rather symmetric geometries such as spheres and larger values corresponding to rod-like and disk-like shapes. In order to determine whether or not an atom is part of a crystallite, we use the adaptive Common Neighbor Analysis method (a-CNA), as implemented in the ovito software package~\cite{Stukowski2009Dec,Stukowski2012May,Honeycutt2002May}.

\section{Results}

\subsection{Chemical ordering and size distribution}

\begin{figure}
\includegraphics[width=\columnwidth]{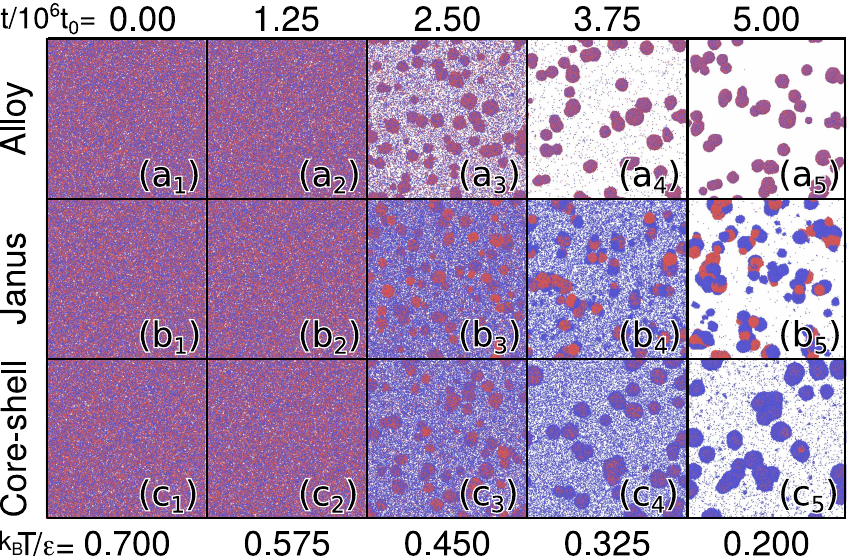}
\caption{Temporal evolution during the temperature quench with $\Delta t=5\times 10^6 t_0$. Time and temperature are given in units of $10^6t_0$ and of $\epsilon$, respectively. In the snapshots, blue and red dots represent A and B atoms. The inert gas atoms are not shown for better clarity.}
\label{ImgVsTime}
\end{figure}

\begin{figure}
\includegraphics[width=\columnwidth]{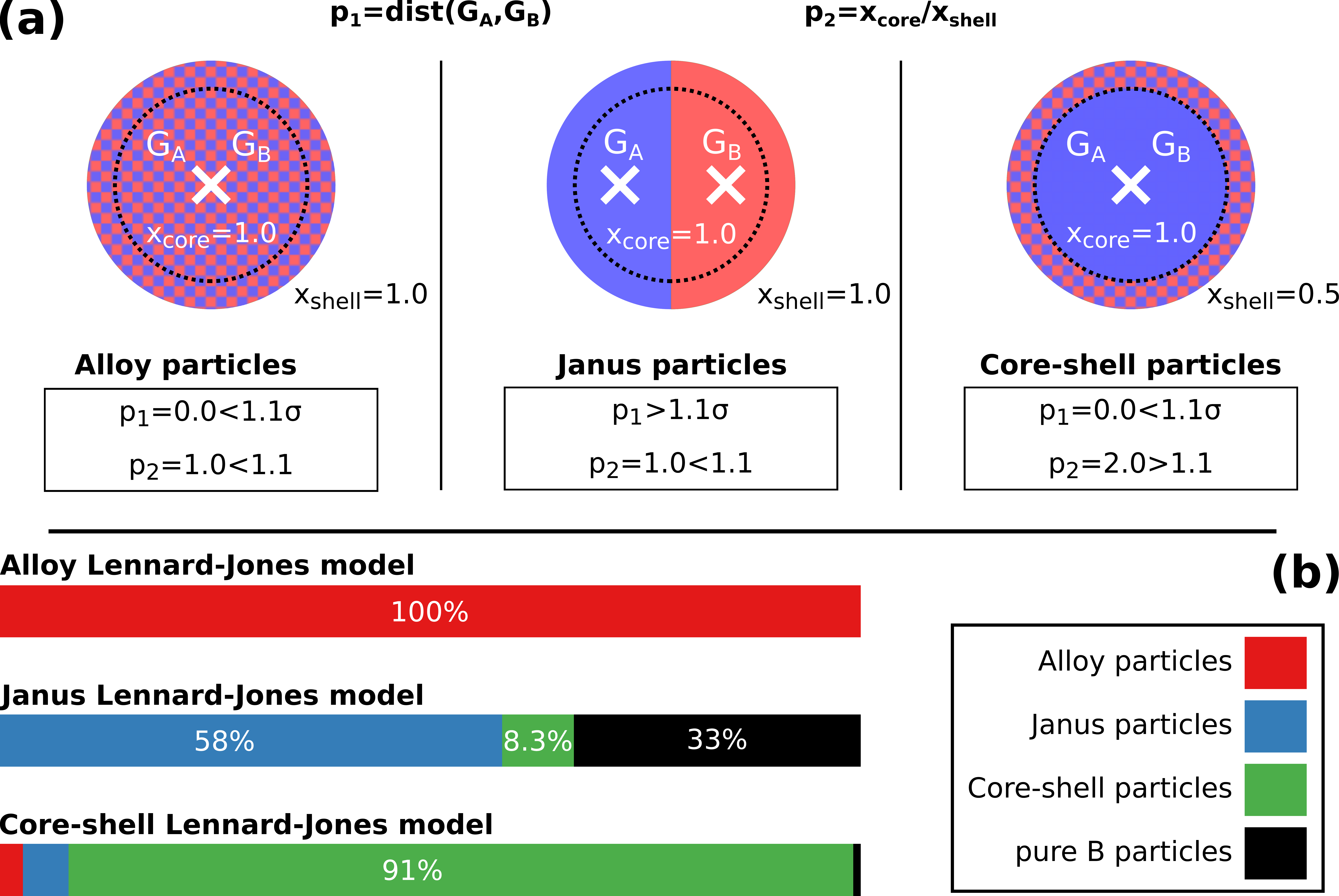}
\caption{(a) Illustration of the order parameters $p_1$ and $p_2$. (b) Chemical ordering for the final NP population obtained at the quench duration of $5 \times 10^6t_0$.} 
\label{composition}
\end{figure}

\begin{figure}[h!]
\includegraphics[width=\columnwidth]{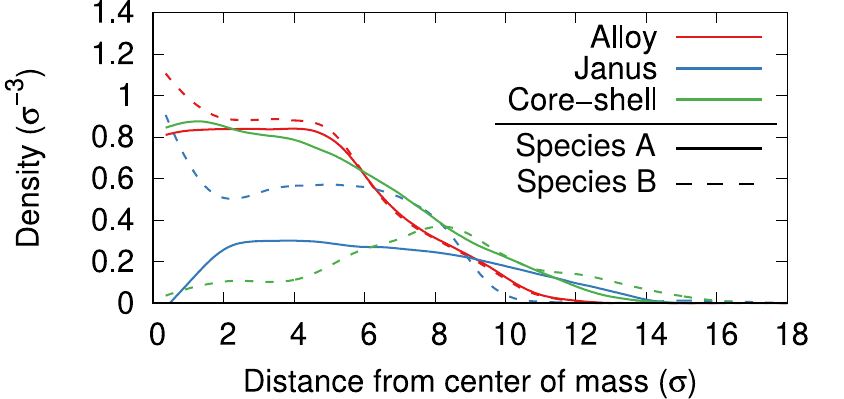}
\caption{Atomic number densities of atoms of species A (solid lines) and species B (dashed lines) as a function of distance from the center of mass of the NPs. Red, blue and green curves indicate alloy, Janus, and core-shell systems, respectively. The data correspond to the largest NP obtained at the end of the quenching simulations with $\Delta t=5.0\times 10^6t_0$.}
\label{rdf}
\end{figure}

In the following two sections, we focus on results obtained at an intermediate quench duration of $\Delta t=5\times10^6 t_0$, i.e., 4.8\,ns in physical units. The final snapshots of Fig.~\ref{ImgVsTime} confirm that alloy, Janus, and core-shell were indeed obtained. The NPs can be assigned one of the three classes by defining two order parameters: (i) $p_1$ is the distance between the center of masses $G$ of atoms of species A and B, and (ii) $p_2$ the ratio of the fraction of atoms of species A denoted $x$ in the core (atoms within the gyration radius) and the shell (all other atoms). Fig.~\ref{rdf} shows that the core-shell NPs of this work consist of a core of atoms A and a shell of an alloy of A and B. Therefore, in our case, $p_2$ would equal two for perfect core-shell NPs, whereas it would equal one for ideal alloy or Janus NPs. The definitions of the order parameters are illustrated in Fig.~\ref{composition}(a). The NP is assigned alloy character if both order parameters are less than thresholds set to 1.1~$\sigma$ and 1.1 respectively, if not the NP is of Janus ($p_1 > p_2$) or core-shell ($p_1 \leq p_2$) type.
For this analysis, we consider only NPs with more than 100 atoms. The majority of the NPs have the desired chemical ordering for all of the three Lennard-Jones models. This is shown in Fig.~\ref{composition}(b) for $\Delta t=5\times10^6 t_0$, but holds true also for all the other quenching rates (data not shown). However, we note that in some cases, a significant number of NPs solely made up of atoms B, form. As we will see in the following, these NPs are smaller and are formed at the end of the quenching process. Corresponding NPs containing only atoms of species A do not occur in the simulations reported here.

Fig.~\ref{rdf} shows the species-resolved atomic number densities within the largest NP obtained at the end of the simulations as a function of the distance from their respective center of masses. Alloy and Janus NPs ideally show uniform distributions, while in the case of the core-shell NPs, higher concentrations of species A can be found in the center as compared to the shell. However, neither, core nor shell are made up purely of either species A or B. In particular, the shell is composed of almost as many atoms of species B as of species A.

\begin{figure}[h!]
\includegraphics[width=\columnwidth]{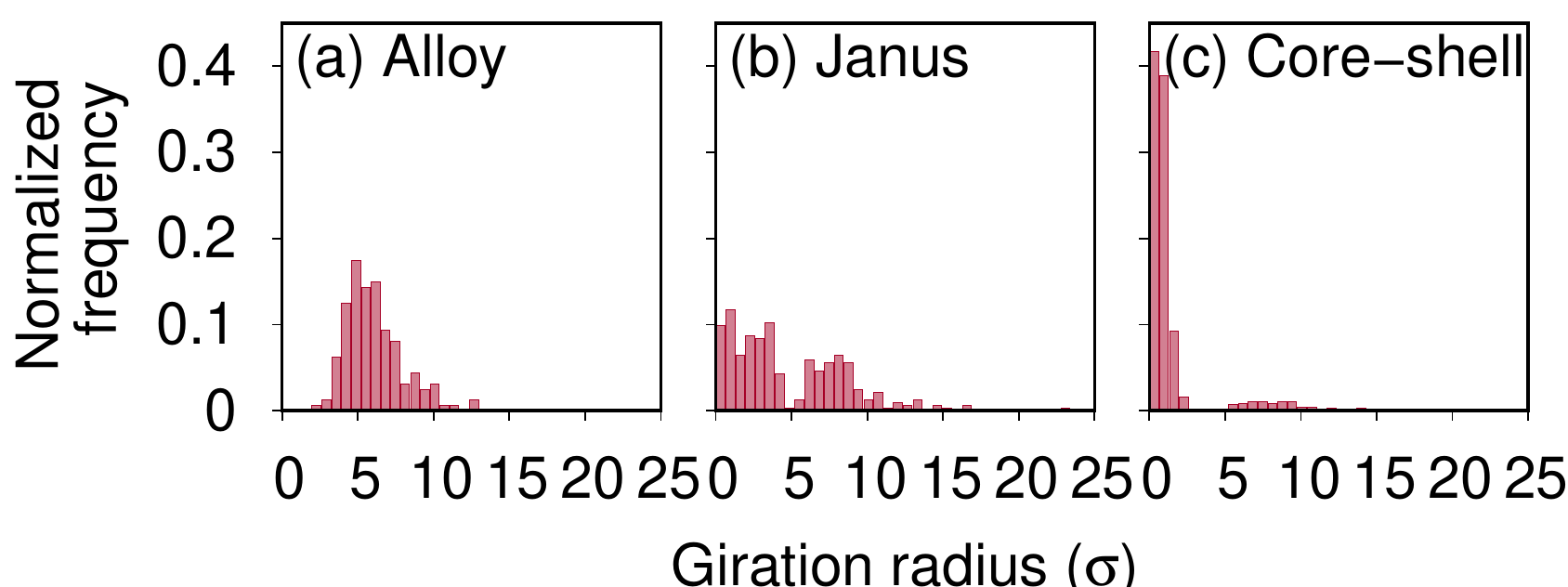}
\caption{Probability distribution of NPs from the final configuration of simulations with a quench duration of 5$\times$10$^6t_0$. Panel (a-c) are concerned with the alloy, Janus, and core-shell systems.}
\label{giration_radius}
\end{figure}

Regarding the size of the obtained NPs, probability distributions of gyration radii are given in Fig.~\ref{giration_radius} for the three system types. The narrowest distribution is achieved within the alloy system, where 59\% of the NPs have a gyration radius in the range of 4.2--6.2 $\sigma$. The situation is qualitatively different for the Janus and core-shell systems: the distribution of the gyration radii is bimodal with relatively broad peaks below and above 5$\sigma$. The small size peak consists in NPs made mostly of atoms of species B that bind more weakly both with one another and atoms of species A.

\subsection{Nucleation and growth mechanisms}

\begin{figure}[h!]
\includegraphics[width=\columnwidth]{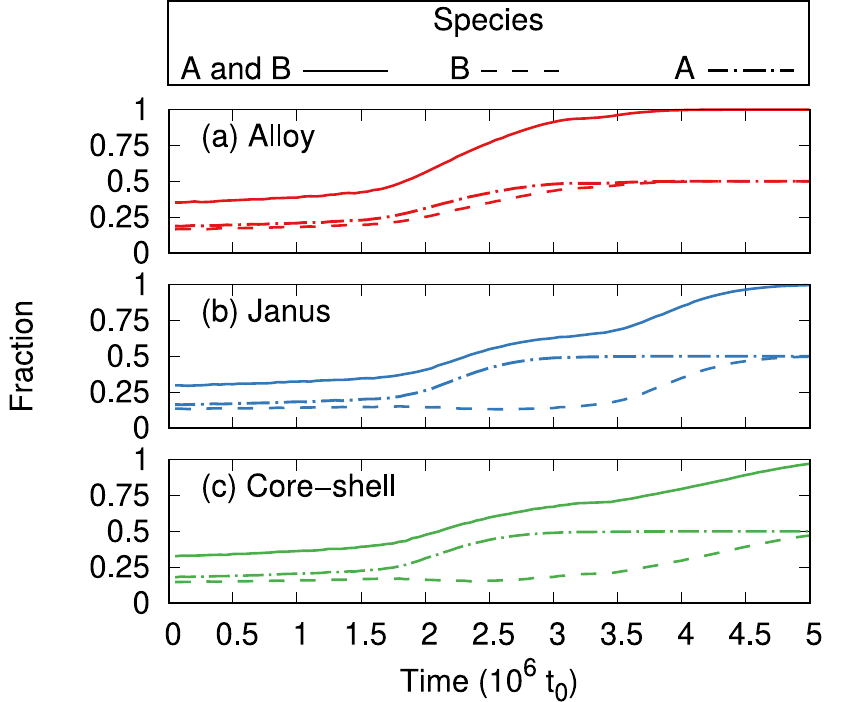}
\caption{Species-resolved fraction of atoms within NPs obtained at a quench duration of $5 \times 10^6t_0$ in the (a) alloy, (b) Janus and (c) core-shell cases.}
\label{fract_evolution}
\end{figure}

In this section, we investigate the temporal evolution i.e., the emergence of the NPs. In Fig.\,\ref{ImgVsTime}, one can roughly observe three stages (1) binary gas, (2) coexistence of lower density gas and droplets, and (3) coexistence of very low-density gas and crystalline NPs with the desired chemical ordering. 
The evolution of the fraction of atoms of species A and B in the forming NPs are shown in Fig.\,\ref{fract_evolution}. While in the alloy case species A and B behave similarly, in the Janus and core-shell cases, two-step mechanisms are observed in which the number of atoms A increases before that of atoms B. 
This shows that the nucleation and growth mechanisms are qualitatively very different for alloy, Janus and core-shell NPs. Inversely, the gas is dominated by monomers of species B during much of the simulation, and especially in the time frame of $2.0-4.0\times$10$^6t_0$ in the case of the Janus and core-shell systems. Droplets condensed from this remaining gas make up the small size peak observed in the size distribution [See Fig.~\ref{giration_radius}].

\begin{figure}[h!]
\includegraphics[width=\columnwidth]{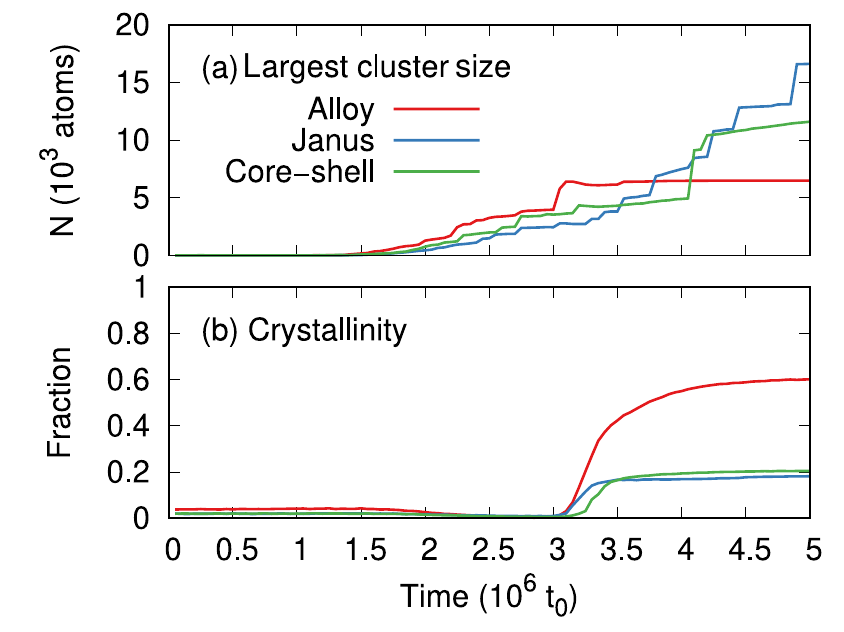}
\caption{Number of atoms within the largest NP and overall fraction of crystalline atoms within NPs as a function of time. The figure shows data corresponding to the alloy, Janus, and core-shell system at a quench duration of 5$\times$10$^6t_0$.}
\label{size_cryst_evolution}
\end{figure}

In order to get a more detailed picture, two parameters are evaluated in time, the size of the largest NP and the fraction of atoms that are found within the crystalline parts of NPs [See Fig.\,\ref{size_cryst_evolution}]. Firstly, the size of the largest NPs increases slowly for $t<2\times 10^6t_0$, which corresponds to nucleation and growth through monomer accretion, and then they grow in a step-wise manner through collisions with other NPs. Secondly, the crystalline structures emerge later at $t>3\times 10^6t_0$, which corresponds to $k_BT<0.4\epsilon$, and then plateau rapidly.

\begin{figure}[h!]
\includegraphics[width=\columnwidth]{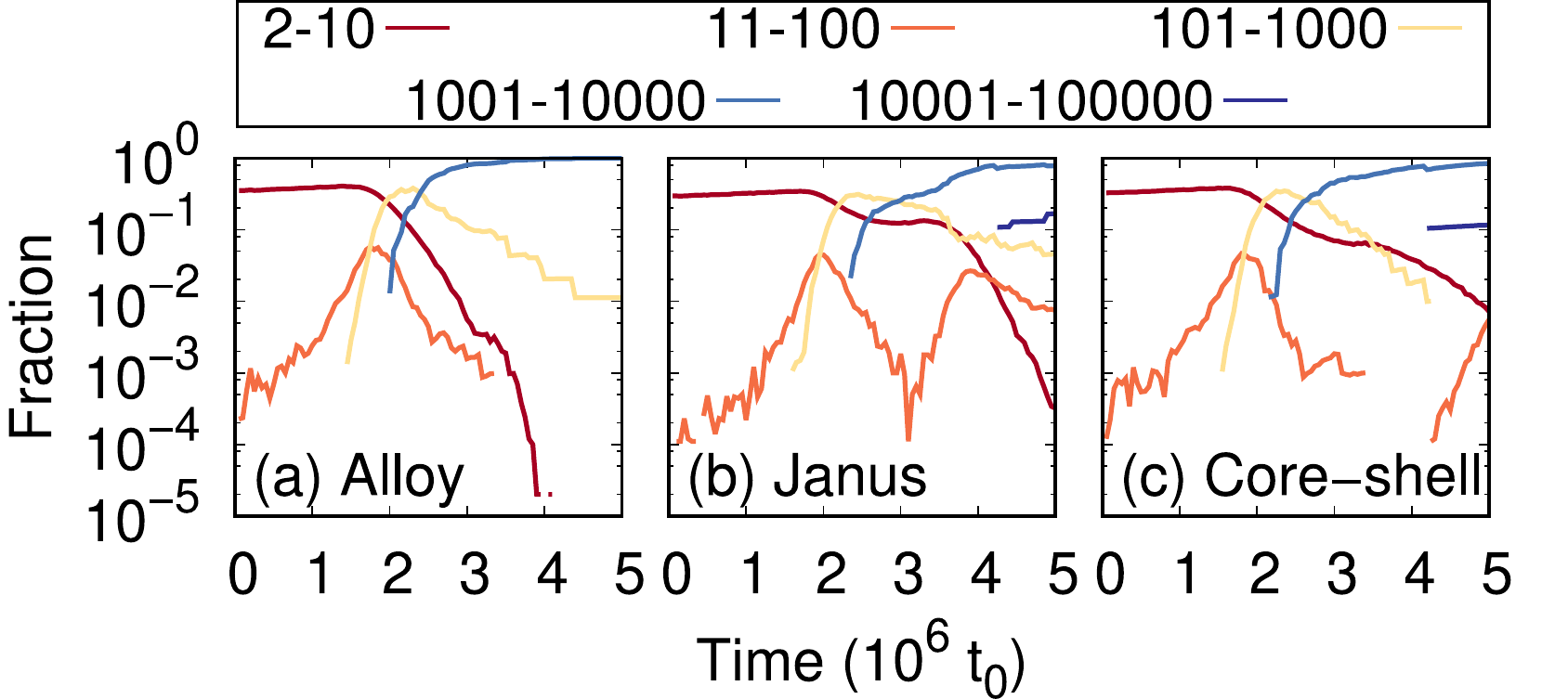}
\caption{Fraction of atoms within NPs of different sizes as a function of time in case of the simulation with a quench duration of 5$\times$10$^6t_0$ for the alloy, Janus, and core-shell systems.} 
\label{size_evolution}
\end{figure}

Not only the size of the largest NPs increases during the quench but also, and more importantly, the entire size distribution evolves. For further analysis, the NPs are grouped by size into five classes: 2-10, 11-100, 101--1,000, 1,001--10,000, and 10,001--100,000 atoms. For the quench duration of 5$\times$10$^6t_0$, Fig.~\ref{size_evolution} shows the abundance of atoms within each of the size classes as a function of time in the case of the alloy, Janus, and core-shell systems. The fraction of atoms within the smallest NPs of up to 10 atoms decreases monotonically with time in the alloy system, which is consistent with results obtained with pure germanium NPs~\cite{Krasnochtchekov2005Jan,Krasnochtchekov2005Oct}. In the case of the Janus and core-shell systems, the curves present local maxima at about the time when significant amounts of atoms of species B condense [See also Fig.~\ref{fract_evolution}]. The amount of atoms within the next larger class of NPs (11-100 atoms) exhibits a maximum at a simulation time of $\approx 2.0\times$10$^6t_0$. While, in the case of the alloy system, this is the only maximum, a secondary lower maximum, which corresponds to the formation of droplets mostly made up of atoms of species B, is found at $\approx 4.0\times$10$^6t_0$ for the Janus NPs. This secondary maximum is not observed for the core-shell NPs either, but the fraction of atoms in NPs of this type and size class increases again at the end of the simulation. Independent of the system type, during a short time window (2.0--2.5$\times$10$^6t_0$), most of the atoms are bound in NPs of 101--1,000 atoms. This particular intermediate size corresponds to the critical nucleus for the gas to liquid transition as observed at approximately 300 atoms with monodispersed Lennard-Jones particles using umbrella sampling~\cite{TenWolde1998Dec}. Later during the simulations, this fraction decreases again at the expense of larger NPs. The amount of atoms within the two largest classes of NPs increases monotonically with time in each of the systems, albeit none of the largest NPs are formed in the case of the alloy system.

\subsection{Tuning structural parameters via the quenching rate}

\begin{figure}[h!]
\includegraphics[width=\columnwidth]{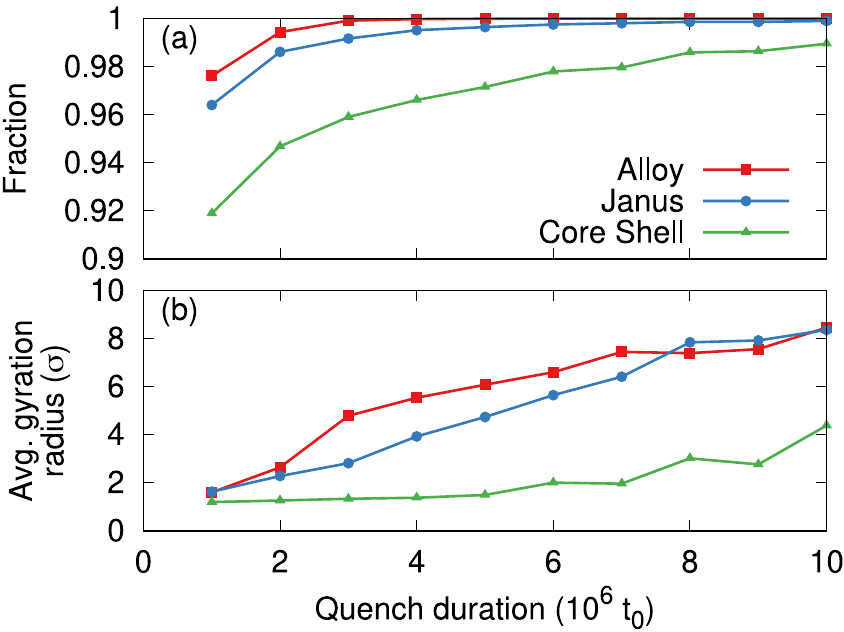}
\caption{Fraction of atoms bound in NPs as well as the average gyration radii of NPs populations as a function of quench duration for the alloy, Janus and core-shell system at the end of the quenching simulation.}
\label{fract2}
\end{figure}

In what follows, we explore how the quenching rate influences the probability distributions of the properties of the NPs. We focus on NP size, shape, crystallinity, and composition at the end of the quenching simulations. The fraction of atoms within NPs increases with quench duration [See Fig.~\ref{fract2}]. The same tendency was also noted in simulations where the inert gas pressure was varied~\cite{Kesala2007May,Krasnochtchekov2005Jan,Krasnochtchekov2005Oct}. The differences between the three system types -- alloy, Janus, and core-shell -- are related to their different condensation temperatures as well as kinetic effects. While in the alloy case, atoms of each type may condense at the NP surface, condensation is limited to one side of the NP for either species in the case of the Janus NPs and one species in the case of the core-shell NPs. In either case, however, most ($>$ 90\%) of the atoms of the system atoms are bound in NPs. As can be appreciated from Fig.~\ref{fract2}(b), not only the fraction of atoms in NPs, also the average size of the NPs increases naturally with the quench duration. This increase in size is particularly pronounced with the alloy and Janus systems. As stated earlier, the size distribution of the Janus and core-shell NPs is bimodal, which translates here in these cases to slightly lower average gyration radii.

\begin{figure}[h!]
\includegraphics[width=\columnwidth]{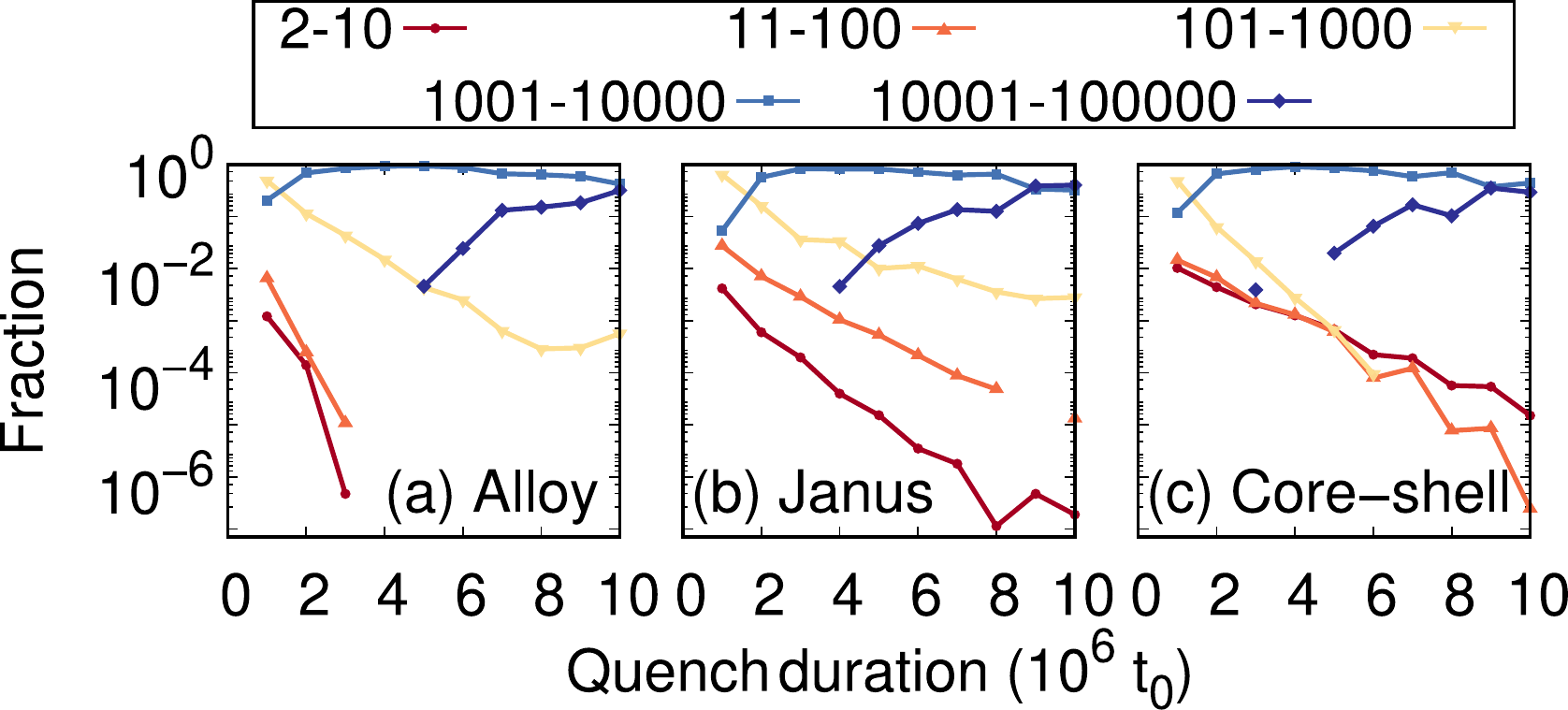}
\caption{Fraction of atoms within NPs of different sizes as a function of the quench duration for the alloy, Janus, and core-shell systems.}
\label{particle_size}
\end{figure}

The abundance of NPs within the different size classes as a function of quench duration for the three system types considered in this work is shown in Fig.~\ref{particle_size}. The fraction of atoms found within the three classes of smallest NP size (up to 1,000 atoms) decreases exponentially (mostly straight lines on the logarithmic plot) with quench duration. While the rate of decrease becomes smaller with NP size in the case of the alloy and Janus systems, it becomes more important for the larger core-shell NPs. Considering the fact that, per NP, the second and third smallest class of NPs contain respectively one and two order of magnitude more atoms than the smallest class of NPs, less of these larger NPs are generated in most cases. The majority of atoms are part of NPs of medium size (1,001--10,000 atoms) at the end of almost all the simulations. Considering the logarithmic scale of Fig.~\ref{particle_size}, the number of atoms within these NPs reaches a clear maximum at a quench duration of about $4\times 10^6t_0$, where they essentially make up the entire system. At larger quench durations, the number of atoms in the largest NPs of up to $1\times 10^5$ atoms increases exponentially and becomes comparable to the atoms within the second largest class of NPs at the largest quench durations ($0.9-1.0\times 10^7t_0$) studied here. Since the simulations used here contain $1\times 10^5$ atoms, we cannot observe larger NPs than that. We speculate, however, that their number would be very limited in the range of quench durations discussed here. Their abundance would become more important at larger quench durations at the expense of the two largest classes of NPs shown in Fig.~\ref{particle_size}.

\begin{figure}[h!]
\includegraphics[width=\columnwidth]{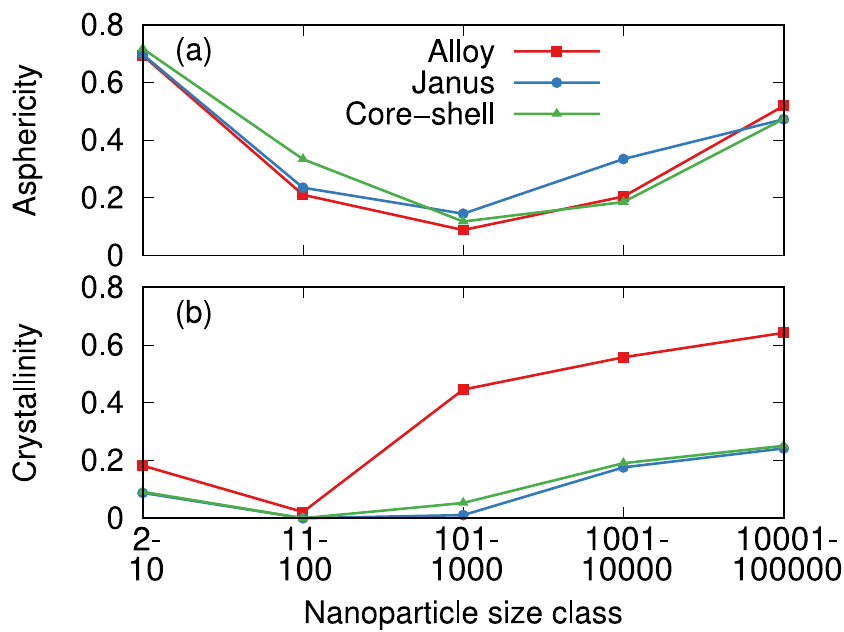}
\caption{Asphericity and crystallinity of NPs of different size classes at the end of the quenching simulations.}
\label{Asphericity_size}
\end{figure}

In the following, we characterize the NPs in terms of their shape, crystallinity, and composition. It turns out that the asphericity of NPs does not depend very much on the quench duration (data not shown), or on the type of system (alloy, Janus, or core-shell). Fig.~\ref{Asphericity_size} shows the average asphericity over all quenching rates as a function of NP size. While the smallest and largest NPs have higher asphericity indexes on average, the most spherical NPs have an intermediate size of 101--1,000 atoms. This size range was already identified as a critical size of the gas-liquid transition when studying the temporal evolution. As such, it suggests that the nucleation core should be spherical in these systems. It is, therefore, instructive to consider the internal structure of the NPs: Fig.~\ref{Asphericity_size} shows the fraction of crystalline atoms within the NPs, averaged over all quenching rates, as well. Surprisingly, the crystallinity is also little dependent on the quench duration (not shown). However, NP size has a significant effect again: The largest NPs have a higher amount of crystalline atoms. This is to be expected, as larger NPs have relatively fewer surface atoms, which are identified as non-crystalline by the a-CNA method.

\begin{figure}[h!]
\includegraphics[width=\columnwidth]{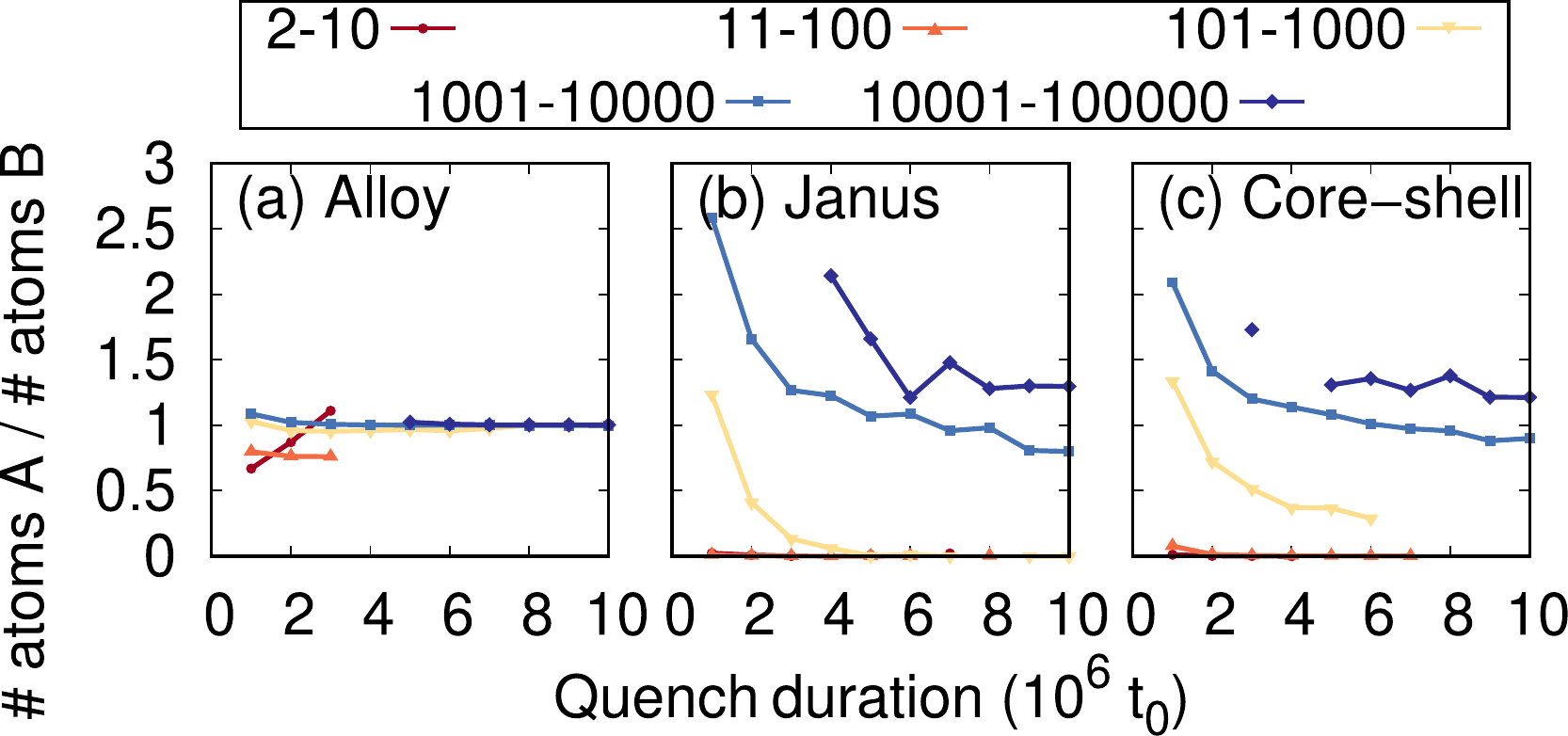}
\caption{Ratio of atoms of species A and B within the final NPs grouped into 5 size classes as a function of quench duration.}
\label{fract_atomtype}
\end{figure}

The chemical composition of the NPs at the end of the quenching simulation is given in Fig.~\ref{fract_atomtype}. In almost all our simulations (except for the smallest alloy NPs at quench durations in the range of 2--3$\times 10^6t_0$), the amount of atoms of species A increases with NP size and decreases with quench duration. The alloy ratio of the two largest NP populations reaches about unity in all three system types at the highest quench durations studied here. In the alloy system, all NP populations have an alloy ratio between 0.65 and 1.15, meaning they always contain a non-negligible number of atoms of either species. In the case of the Janus and core-shell systems, this ratio tends to zero for NPs smaller than 1,000 atoms. In other words, these NPs are composed mostly of atoms of species B. Atoms of species B condense at lower temperatures and thus later in the quenching simulations (compare Fig.~\ref{fract_evolution}). At that time, atoms of species A are already absorbed into other NPs that are then larger at the end of the simulation. Therefore, close to the end of the simulations, there is mostly gas of atoms B left that finally condenses into small, almost monoatomic NPs of species B. Inversely, atoms of species A condense at higher temperatures, and, in the case of the core-shell NPs form the core of the NPs.

\section{Conclusions}

In summary, molecular dynamics simulations were employed in order to study the formation mechanisms occurring in the physical synthesis of binary NPs. First, we confirm that with simple Lennard-Jones interactions, one can obtain NPs with different chemical ordering: alloy, Janus, and core-shell. Analyzing the probability distribution, we show that while alloy NPs display narrow size distributions, Janus and core-shell NPs exhibit bimodal distributions. Our simulations reveal that the nucleation and growth mechanisms depend qualitatively on the system type, with two-step processes occurring in Janus and core-shell systems. A critical cluster size emerges at intermediate time which is reminiscent of a classical nucleation core in equilibrium conditions. Studying the temporal evolution, we go beyond equilibrium simulations. In our case, the phase transitions occur at low free energy barriers and kinetic considerations related to quench durations are more likely to explain the formation mechanisms. The quench duration does not affect the crystallinity nor the asphericity of the NPs but remains an essential parameter to tune NP size and the chemical ordering, especially for Janus and core-shell systems. While our modeling follows the simplest possible approach, it allowed for studying numerous features of the physical synthesis of binary NPs. As such, we hope it will encourage further studies with the same modeling, but even more complex structural characterization using for instance novel machine-learning methodologies~\cite{Fernandez2015Nov}.

\section{Acknowledgement}

JL acknowledges financial support of the Fonds de la Recherche Scientifique - FNRS. Computational resources have been provided by the Consortium des Equipements de Calcul Intensif (CECI) and by the F\'ed\'eration Lyonnaise de Mod\'elisation et Sciences Numériques (FLMSN).

\end{document}